\documentclass[trackchanges,twocolumn]
{aastex701}

\usepackage{subcaption}
\usepackage{graphicx}
\usepackage{diagbox}
\usepackage{booktabs}
\usepackage{amsmath}
\usepackage{makecell}
\usepackage{array}
\usepackage{multirow}
\usepackage{mathrsfs}
\usepackage{CJK}

\newcommand{\msun}{M_\odot}

\newcommand{\zsun}{Z_\odot}

\newcommand{\cc}{{\rm cm}^{-3}}
\newcommand{\Ncc}{{\rm cm}^{-2}}

\newcommand{\msunyr}{M_\odot~{\rm yr}^{-1}}

\newcommand{\pc}{{\rm pc}}

\newcommand{\kms}{{\rm km~s}^{-1}}

\newcommand{\K}{{\rm K}}
\newcommand{\beq}{\begin{equation}}
\newcommand{\eeq}{\end{equation}}

\newcommand{\D}{{\rm d}}

\newcommand{\Ha}{\mathrm{H}\alpha}
\newcommand{\Hb}{\mathrm{H}\beta}
\newcommand{\Hg}{\mathrm{H}\gamma}
\newcommand{\Paa}{\mathrm{Pa}\alpha}
\newcommand{\nH}{n_\mathrm{H}}
\newcommand{\NH}{N_\mathrm{H}}

\begin{document}
\begin{CJK*}{UTF8}{gbsn}

\title{Balmer Transition Signatures from Gas-Enshrouded, Dust-Poor Active Galactic Nuclei}

\author[orcid=0009-0008-9591-7052,sname='Yan']{Zu Yan (晏祖)}
\affiliation{Kavli Institute for Astronomy and Astrophysics, Peking University, Beijing 100871, China}
\affiliation{Department of Astronomy, School of Physics, Peking University, Beijing 100871, China}
\email[show]{yanzu@stu.pku.edu.cn}  

\author[orcid=0000-0001-9840-4959,sname='']{Kohei Inayoshi}
\affiliation{Kavli Institute for Astronomy and Astrophysics, Peking University, Beijing 100871, China}
\email[show]{inayoshi@pku.edu.cn}  

\author[orcid=0009-0005-1831-3042,sname='']{Kejian Chen (陈可鉴)}
\affiliation{Kavli Institute for Astronomy and Astrophysics, Peking University, Beijing 100871, China}
\affiliation{Department of Astronomy, School of Physics, Peking University, Beijing 100871, China}
\email{chenkejian@stu.pku.edu.cn}  

\author[orcid=0009-0002-6578-8110,sname='']{Jingsong Guo (郭京松)}
\affiliation{Department of Astronomy, School of Physics, Peking University, Beijing 100871, China}
\email{2100011602@stu.pku.edu.cn}

\begin{abstract}
Little red dots (LRDs), a population of active galactic nuclei (AGNs) recently discovered by JWST, show 
distinctive Balmer-transition features, including prominent Balmer absorption, pronounced Balmer breaks, and 
large equivalent widths of broad $\Ha$ emission, all of which indicate the presence of dense gas surrounding 
their central black holes.
A further key property of LRDs is their large Balmer decrements with broad $\Ha/\Hb$ line-flux ratios far exceeding 
the Case~B recombination value.
These ratios of $\Ha/\Hb>3$ have often been interpreted as evidence for heavy dust extinction ($A_V\gtrsim 3$ mag), 
however such dust would inevitably produce strong near-to-mid infrared re-emission that is hardly seen in JWST/MIRI observations.
To investigate the physical origin of these observed Balmer features, we perform radiation transfer calculations through dust-free, dense gas.
We show that the observed large Balmer decrements ($\Ha/\Hb$ and $\Ha/\Hg$) naturally arise from Balmer resonance scattering without invoking dust.
At sufficiently high densities ($\nH \gtrsim 10^{{8}-{10}}~\cc$), the elevated multiple Balmer-line ratios converge to 
values that closely mimic dust reddening, explaining why LRD spectra resemble obscured AGNs.
Furthermore, when the Balmer break and broad Balmer lines originate in the same dense gas,
their strengths are physically linked, allowing us to constrain the density structure and infer 
a low broad-line region gas mass of $\sim \mathcal{O}(10~\msun)$.
Such a small gas reservoir would be enriched by even a single supernova, implying that LRDs with observed 
low-metallicity signatures likely experienced minimal star formation in their nuclei.
\end{abstract}


\section{Introduction} 
\end{CJK*}

The James Webb Space Telescope (JWST) has uncovered a population of compact red objects at $z>4$, 
known as little red dots \citep[LRDs;][]{Kocevski2023,Kokorev_2023_aLRDwNIRSpec,Matthee2024,Labbe_2025_LRD_ALMA}.
These sources show a set of distinctive observational signatures: V-shaped UV-optical spectral 
energy distributions (SEDs), broad Balmer emission lines, and a lack of X-ray emission \citep{Barro2024,Greene2024,Setton2024,Maiolino2025}.
Although their physical nature remains under debate, a growing number of evidence suggests that 
LRDs are powered by accreting massive black holes (BHs) and may represent a key early phase of BH growth \citep{Inayoshi&Ho2025}.

JWST spectroscopy reveals that the Balmer-line properties of LRDs differ from those of typical AGNs.
First, prominent absorption features are often associated with broad Balmer emission lines \citep{Matthee2024,Lin2024,Kocevski2025}, while nearby AGNs rarely show such signatures \citep{Aoki2006,Shi2016,Schulze2018}.
Second, LRDs exhibit deep spectral breaks near $4000~{\rm \AA}$ that in some cases cannot be explained by stellar continua of 
evolved populations \citep{Naidu2025,deGraaff25,deGraaff25b,Taylor25}.
Third, a large fraction of JWST-identified high-redshift AGNs (including LRDs) have strong, broad H$\alpha$ emission with rest-frame equivalent widths (EW) 
$\sim 3$ times larger than in normal AGNs \citep{Maiolino2025}.
These Balmer-line characteristics point to a unique nuclear environment of LRDs, where the broad-line region (BLR) is embedded in dense gas
absorbers and line-emitters with a high covering fraction \citep{Inayoshi&Maiolino2025,Maiolino2025}.

Another key signature of LRDs is their large Balmer decrements of H$\alpha$/H$\beta$ flux ratios \citep{Rusakov2025,Nikopoulos2025,2025arXiv251000101D,Torralba2025}
exceeding the value expected from the Case~B recombination of $\simeq 2.86$ \citep{Osterbrock_1974} or its partially optically thick variants of 
$\simeq 3.1$ \citep{Osterbrock&Ferland2006}.
Such elevated $\Ha/\Hb$ values are commonly interpreted as a result of dust extinction (\citealt{Calzetti1994,Dong2008,Brooks2025}; but see also
the argument by \citealt{Son2025})
with $A_V \sim 3 ~{\rm mag}$ in LRDs, consistent with their red optical continua.
Using an LRD sample of high-S/N JWST/NIRSpec spectra, \cite{Nikopoulos2025} recently found that broad Balmer-line ratios 
largely depart from Case B, whereas narrow lines remain consistent.
The broad-line ratios (H$\alpha$/H$\beta$ and H$\alpha$/H$\gamma$) correlate in a manner compatible with dust reddening at $A_V \simeq 1-8 ~{\rm mag}$.
However, this interpretation leads to a severe problem. 
Such large $A_V$ values would inevitably produce near-to-mid infrared (NIR-MIR) dust re-emission
that exceeds JWST/MIRI flux detections and upper limits \citep{Setton25,Akins_2025,Williams24,PerezGonzalez2024,Leung2025}.
To satisfy the MIRI (and ALMA) constraints, the extinction must be as low as $A_V \lesssim 1-1.5 ~{\rm mag}$ 
\citep{chen2025dustbudgetcrisislittle}, implying dust masses lower than $\lesssim 10^6~\msun$ \citep{Casey2025,chen2025dustbudgetcrisislittle}.

This tension indicates that the elevated Balmer decrements cannot be explained only by dust reddening. 
Instead, such trends naturally arise in dense, dust-poor gas, where hydrogen collisional excitation and optical-depth effects substantially 
reshape the Balmer line ratios \citep{Baldwin1975,Netzer1975,Krolik&McKee1978}.
In particular, resonance scattering can convert H$\beta$ photons into Pa$\alpha$ and H$\alpha$, enhancing the Balmer decrement without requiring dust
\citep{Krolik&McKee1978,Chang2025}.
If the properties of Balmer absorption, break, and decrement can all be attributed to high-density gas, 
the observations support the BH-envelope model of LRDs, in which an accreting BH is enshrouded within a dense, optically thick gas envelope 
that naturally produces the red optical continua and flat NIR spectra \citep{Inayoshi_2025b,Kido_2025,Liu_2025b,Begelman_2025,X.Lin_2025}.

In this paper, we examine whether dense gas clouds can self-consistently reproduce the observed Balmer-line properties of LRDs,
utilizing CLOUDY simulations.
We find that a large offset of multiple Balmer line ratios (H$\alpha$/H$\beta$ and H$\gamma$/H$\alpha$) from Case~B values, 
which appears to align with
dust reddening scenarios, can be reproduced by dense gas clouds with volume densities of $\nH \simeq 10^{9-11}~\cc$ 
and column densities of $\NH \simeq 10^{22-24}~\Ncc$ without relying on dust.
We further explore a potential correlation between broad Balmer-line ratios and Balmer-break strength when both arise from the same dense clouds,
enabling joint constraints on the physical conditions of the envelope.

\section{Methods} \label{sec:methods} 

To quantify the SED of an incident flux passing through 
a single gas cloud, we make use of the CLOUDY code (C23, \citealt{2023RNAAS...7..246G})
that performs radiation transfer calculations for both continua and lines, coupled with non-local thermodynamic 
equilibrium (non-LTE) modeling of atomic hydrogen level populations.
The incident radiation source is assumed to be an AGN consisting of an accretion disk and non-thermal emission, 
whose SED is approximated by
\begin{equation}
    f_{\nu} \propto \max \left[ \nu^{\alpha_{\mathrm{uv}}} e^{-h\nu 
    / k_{\mathrm{B}} T_{\mathrm{bb}}},\  r_{\mathrm{x}} 
    \nu^{\alpha_{\mathrm{x}}} \right]\,,
\end{equation}
\citep{Inayoshi&Maiolino2025}, where $T_{\rm bb}=10^5~\K$ is the disk characteristic temperature measured at a distance of $\sim 10$ Schwarzschild radii from the central BH with 
a mass of $M_\bullet \simeq 10^{7-8}~\msun$ accreting at the Eddington rate \citep{Novikov&Thorne1973}. 
The UV and X-ray spectral indices are $\alpha_{\mathrm{uv}} = -0.5$ and $\alpha_{\mathrm{x}} = -1.5$.
The normalization of $r_{\mathrm{x}}$ is adjusted so that the spectral slope between 2500 \AA\ and 2 
keV becomes $\alpha_{\mathrm{ox}} = -1.5$ \citep{Lusso_2020}. 
The value of $\alpha_{\mathrm{uv}} = -0.5$ is consistent with that of the low-redshift composite quasar SED \citep{Vanden2001}. 
We determine the flux density normalization by setting the ionization parameter, $U\equiv\Phi_0/(n_{\rm H}c)$ to $ -2 \leq \log U \leq -1 $, 
where $\Phi_0$ is the ionizing photon number flux, 
$n_{\rm H}$ is the number density of hydrogen nuclei, 
and $c$ is the speed of light. 
In this study, we adopt $ \log U = -1.5 $ as the fiducial value.

In our CLOUDY calculations, the incident radiation spectrum is {\it not} assumed to be thermal emission from a BH-envelope
with a surface temperature of $\simeq 5000~\K$. 
Instead, we adopt a harder AGN spectral shape for two reasons:
(1) the observed spectra of LRDs result from gas attenuation and continuum/line emission reprocessed from intrinsic AGN radiation,
and (2) we aim to explore a more general AGN + dense gas configuration beyond the specific case of LRDs, whose nuclear 
properties remain uncertain.

\begin{figure*}[t]
\centering
\includegraphics[width=89mm]{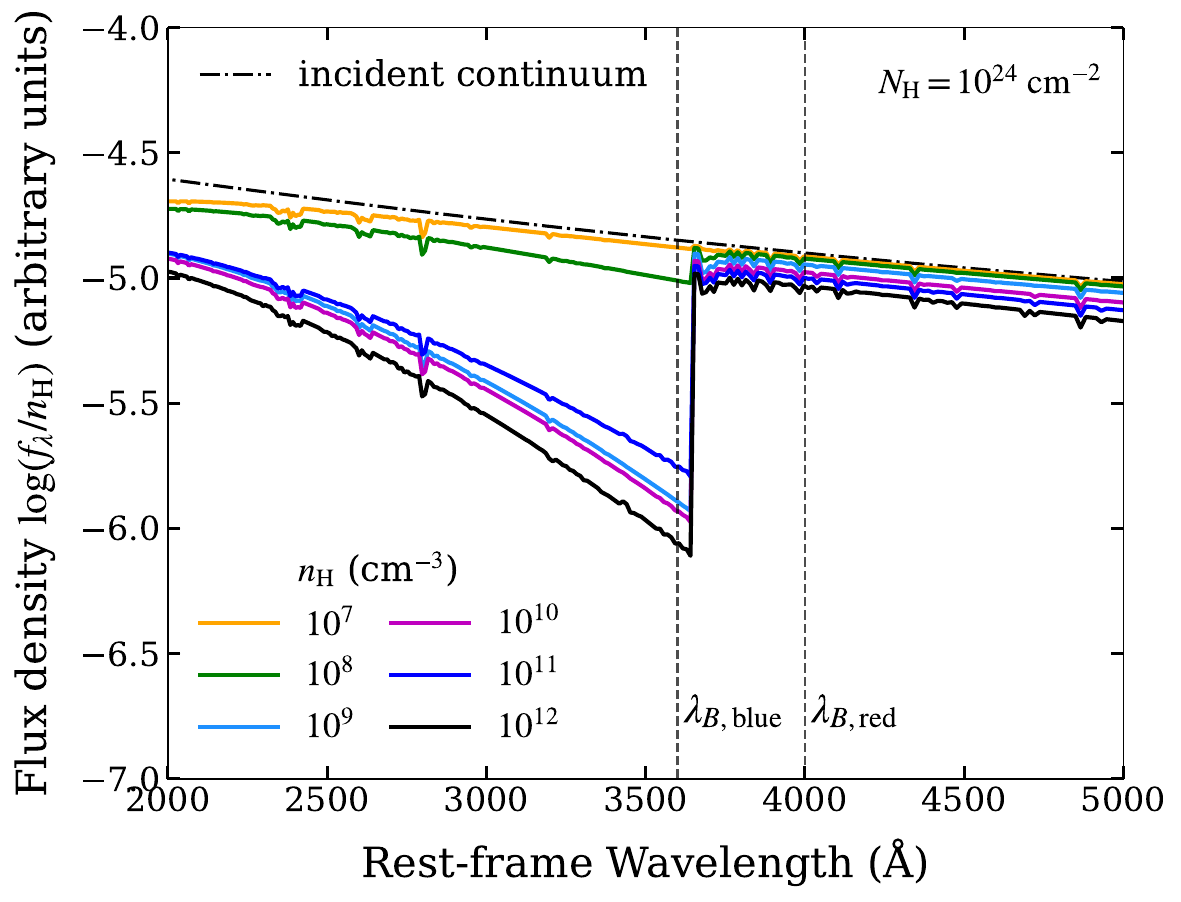}\hspace{1mm}
\includegraphics[width=89mm]{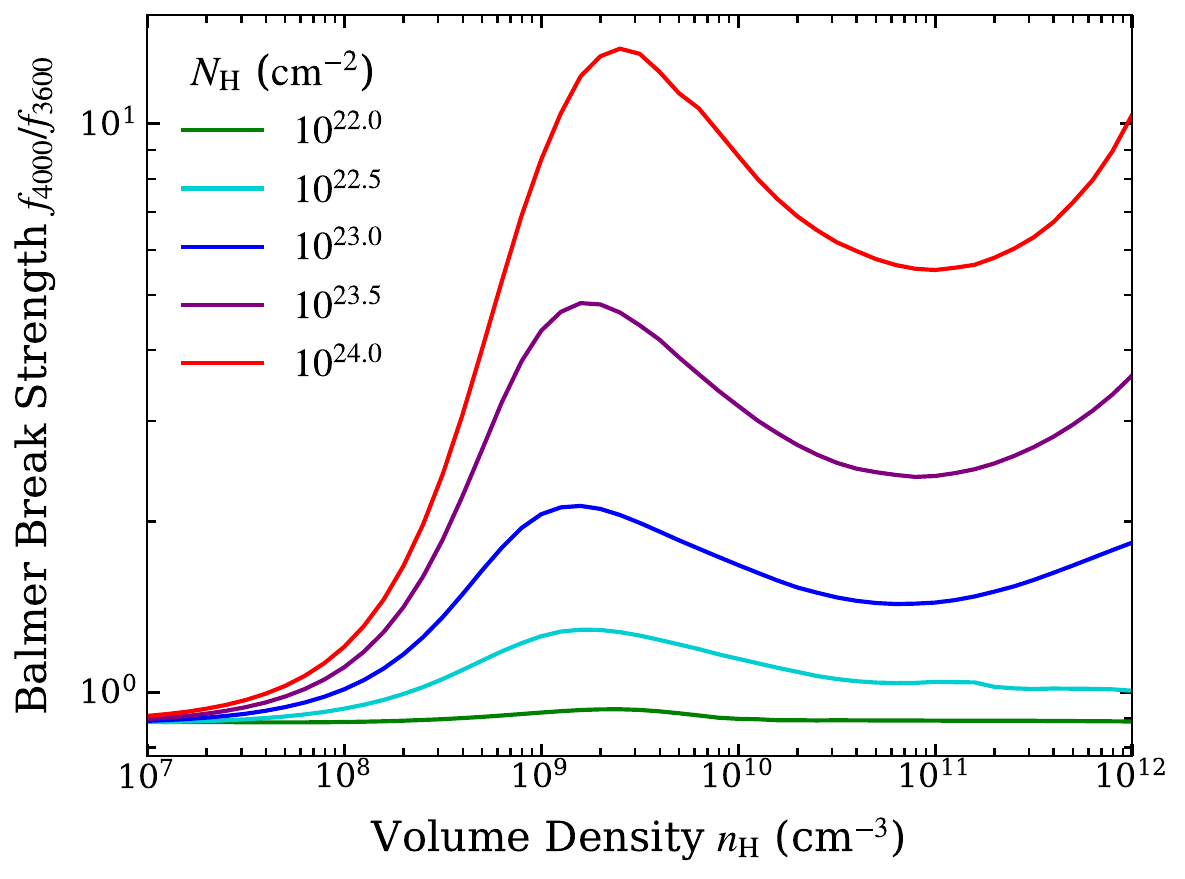}
\vspace{-5mm}
    \caption{
         Left: Transmitted AGN SEDs attenuated by a dust-free gas slab with a gas metallicity $0.1~\zsun$. Nebular emission from the cloud is not included. All spectra correspond to a fixed column density of $N_{\mathrm{H}}=10^{24}\,\mathrm{cm}^{-2}$ and hydrogen volume densities ranging from $\nH=10^{7}$ to $10^{12}\,\cc$. Two vertical lines mark the wavelengths $\lambda_{\mathrm{B,blue}}=3600\, $\AA$ $ and $\lambda_{\mathrm{B,red}}=4000\, $\AA$ $ used to measure the Balmer break strength. The incident AGN continuum is shown with the dashed-dotted curve for comparison. Right: Balmer break strength defined by $f_{\lambda_{\rm B,red}}/f_{\lambda_{\rm B,blue}}$ as a function of hydrogen volume density for column densities $N_{\mathrm{H}}=10^{22}-10^{24}\,\mathrm{cm}^{-2}$.
         The Balmer break strength rapidly increases within $n_{\rm H}\simeq 10^{8-9}~\cc$, reflecting the enhanced population 
         of atomic hydrogen in the $n=2$ level.
    }
    \vspace{3mm}
    \label{fig:SED_BB_nH}
\end{figure*}

We adopt a plane-parallel geometry of a gas slab, assuming that its cross sections are sufficiently small.
The transmitted continuum refers to the incident radiation flux that passes through the cloud, excluding any diffuse nebular emission 
produced within the cloud itself.
The cloud is assumed to fully cover the continuum source along our line of sight.
In the following analysis, we measure the Balmer break strength from the transmitted SED and the Balmer decrement from the nebular component, respectively.
The total SED is calculated by combining the transmitted and nebular components with 
the nebular contribution scaled by a covering fraction $C$, where $C=1$ corresponds to $4\pi$ coverage \footnote{In a plane-parallel geometry,
the intensity of an emitted cloud spectrum depends only weakly on the input covering factor $\tilde C$ specified in the CLOUDY setup.
Note that $\tilde C$ differs from the covering fraction $C$ defined in the main text.
The transmitted SED is entirely independent of $\tilde C$, and the Balmer decrement derived from the nebular emission varies 
by only $\lesssim 2.6\%$ 
across $0.01\leq \tilde C \leq 0.9$.
The primary effect of $\tilde C$ is a modest increase of the nebular emission intensity with a Balmer jump feature.
For our fiducial models, we adopt $\tilde C = 0.5$.}.
It is worth noting that the Balmer break strength depends weakly on the nebular contribution as its Balmer jump feature
partially fills the break and reduces its apparent depth in the total SED, whereas the Balmer decrement value is independent 
of the continuum spectra modeling.

\begin{figure*}[t]
\includegraphics[width=88mm]{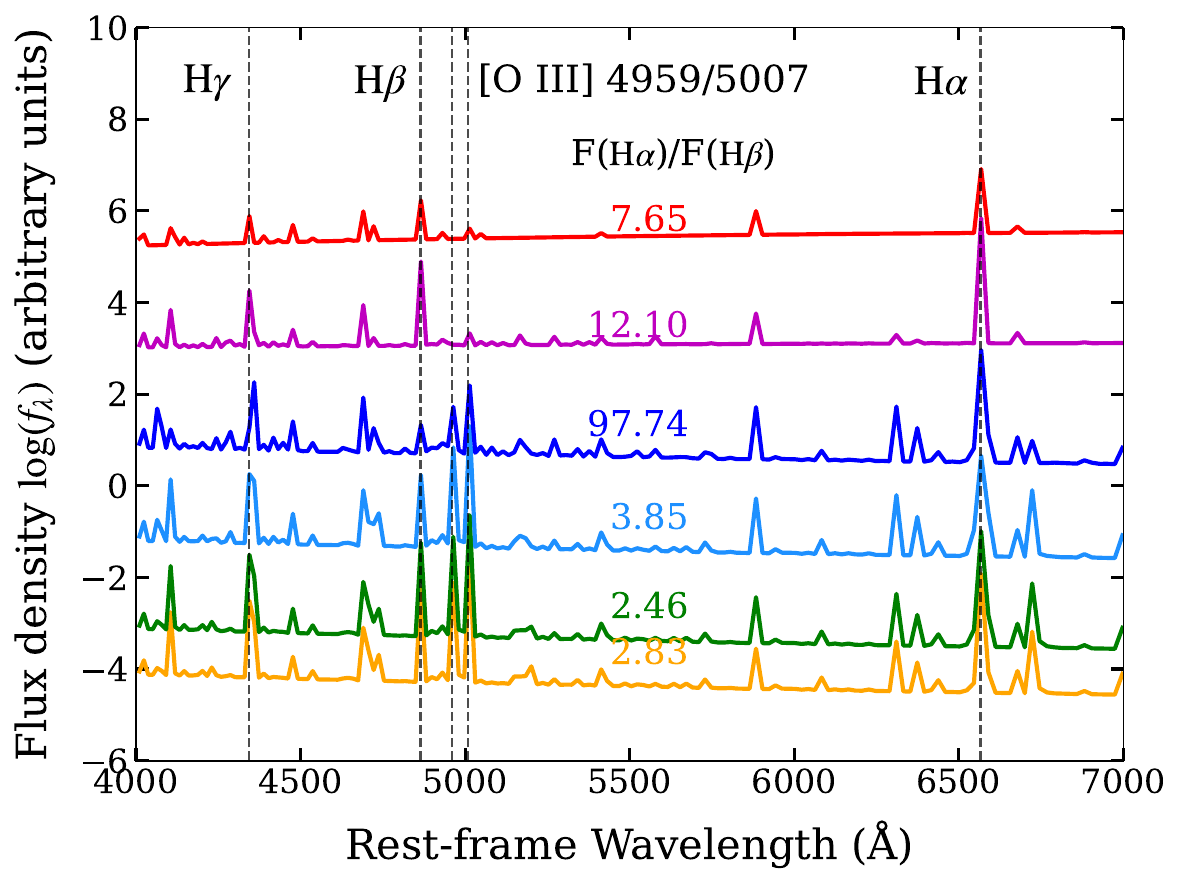}  \hspace{1mm}
\includegraphics[width=88mm]{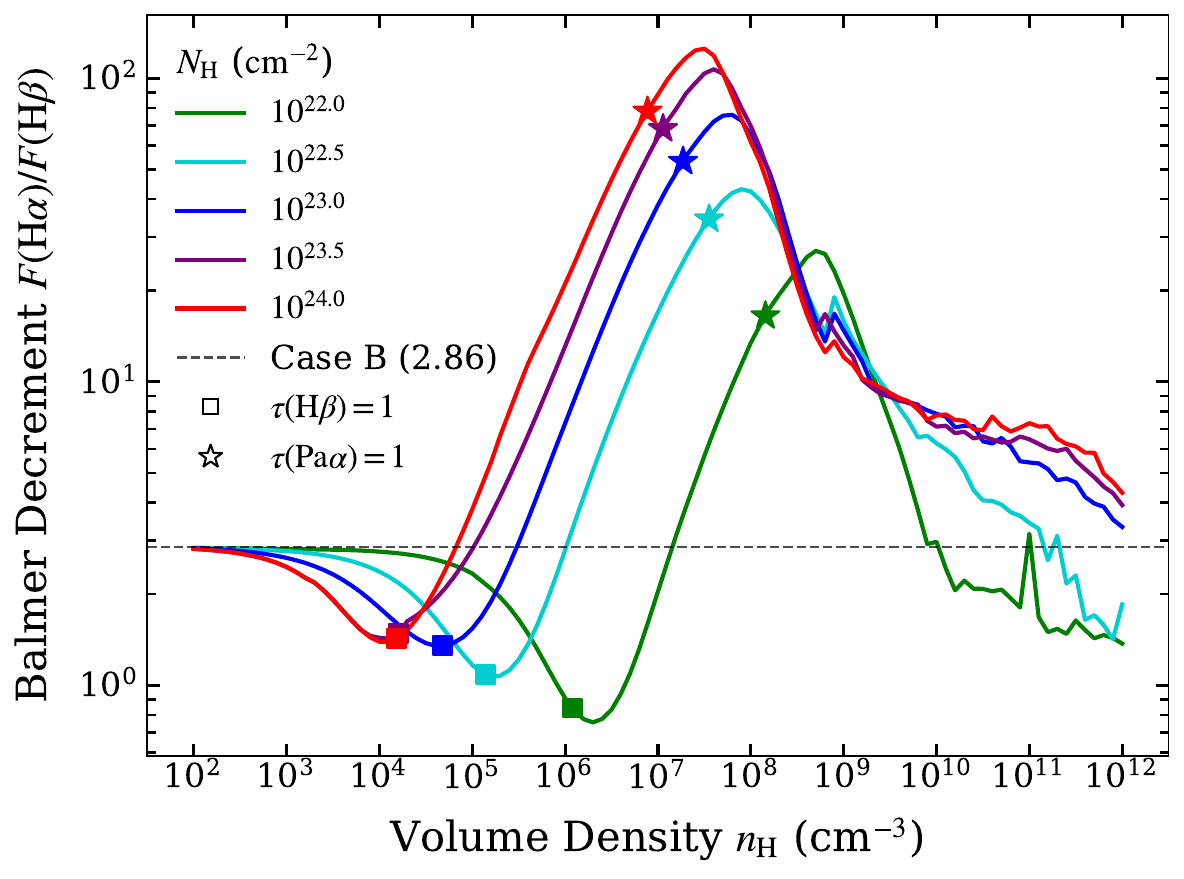}
\vspace{-1mm}
    \caption{
         Left: Emitted continuum and line spectra for a hydrogen column density of $N_{\mathrm{H}}=10^{24}\,\mathrm{cm}^{-2}$. From the bottom to the top, each curve represents the case with hydrogen volume density of $n_{\rm H}=10^2$, $10^3$, $10^5$, $10^7$, $10^9$, and $10^{11}~\cc$. Vertical dashed lines mark the wavelengths of the Balmer lines ($\lambda_{\mathrm{H}\alpha}=6563\, $\AA$ $,\, $\lambda_{\mathrm{H}\beta}=4861\, $\AA$ $ and $\lambda_{\mathrm{H}\gamma}=4341\, $\AA$ $) and the [\ion{O}{3}]~$\lambda\lambda4959,5007$ emission lines. The numbers adjacent to curves of the same color indicate the Balmer decrement, $F(\mathrm{H}\alpha)/F(\mathrm{H}\beta)$, measured directly from the corresponding SEDs. Right: Balmer decrement as a function of hydrogen volume density $n_{\mathrm{H}}$ for different hydrogen column densities $N_{\mathrm{H}}=10^{22}-10^{24}\,\mathrm{cm}^{-2}$. 
        The horizontal dashed line marks the Case~B recombination value of 2.86. Square and star symbols denote the characteristic densities at which H$\beta$ and Pa$\alpha$ become optically thick, respectively.
    }
    \label{fig:SED_BD_nH}
    \vspace{3mm}
\end{figure*}

CLOUDY calculations also generate various outputs of physical quantities that are useful for diagnosing the thermal state of the absorbing and emitting gas clouds. 
These include profiles of the gas temperature, the population densities of atomic hydrogen in energy levels 
with principal quantum number $n(=1,2,\ldots)$ and orbital quantum number $l=(s$, $p$, $d,\dots$), and 
the optical depths of individual bound-bound transitions (e.g., ${\rm H}\alpha$, ${\rm H} \beta$, and ${\rm Pa} \alpha$).
Such information allows us to investigate how dense gas affects the SED properties, specifically the Balmer decrement 
\citep{Netzer1975,Krolik&McKee1978}, under conditions free from dust extinction.

Our fiducial model assumes a metallicity of $0.1~\zsun $, representing low-metallicity environments expected 
for high-redshift AGN and LRDs \citep{Trefoloni2025,Venturi2024,Mazzolari2024}. 
We have checked that both the Balmer break strength and Balmer decrement vary less than $\sim15\%$ depending on the metallicity within the range $0.01\leq Z/Z_{\odot}\leq1$.
In this work, we exclude the effects of dust extinction to isolate and demonstrate how high Balmer decrements 
are reproduced under different physical conditions of dense gas clouds. 
The hydrogen number density is changed over a wide range of $ 10^{2} \leq n_{\rm H} / \cc \leq 10^{12} $ 
in steps of $\Delta \log (n_{\rm H}/\cc)=0.1$, and the CLOUDY calculations are terminated when the total hydrogen column density reaches a given value $\NH$, ranging over $\NH = 10^{22}-10^{24}~\Ncc$ with increments of 
$\Delta \log (N_{\rm H}/\Ncc)=0.5$.

\section{Results} \label{sec:results}

\subsection{Spectral characteristics} \label{subsec:AGN SEDs}

The left panel of Figure~\ref{fig:SED_BB_nH} presents the transmitted AGN SEDs for a fixed column density of
$N_{\rm H}=10^{24}~\Ncc$ and gas densities in the range of $ 10^{7} \leq n_{\rm H} / \cc \leq 10^{12} $. 
At $n_{\rm H}\lesssim 10^7~\cc$, the smooth blue continuum of the incident SED remains nearly unchanged.
Once the density exceeds $n_{\rm H}=10^8~\cc$, the continuum emission at shorter wavelengths of the Balmer limit
($\lambda_{\rm B,lim}=3646~{\rm \AA}$) is strongly suppressed, and the discontinuity across the Balmer limit peaks at 
$n_{\rm H}\simeq 10^9~\cc$ \citep[see also][]{Inayoshi&Maiolino2025,Ji2025}.

The right panel of Figure~\ref{fig:SED_BB_nH} shows the Balmer break strength
as a function of volume density for five column densities ($10^{22}\leq N_{\rm H}/\Ncc \leq 10^{24}$). 
We define the Balmer break strength as $f_{\lambda_{\rm B,red}}/f_{\lambda_{\rm B,blue}}$, using continuum flux densities at 
$\lambda_{\rm B,red}=4000~{\rm \AA}$ and $\lambda_{\rm B,blue}=3600~{\rm \AA}$.
As noted above, the break strength increases with density, reaches a maximum at $n_{\rm H}\simeq 10^{9}~\cc$,
and remains high toward higher densities.
The emergence of a deep Balmer break depends mainly on the gas volume density and only weakly on the column density.
In the low-density limit, the break strength converges to the minimum value set by the intrinsic spectral slope, not by gas attenuation.
At a fixed volume density, the break becomes more prominent as the column density increases, due to a larger population of atomic hydrogen 
in the $n=2$ level.
For high column densities ($N_{\rm H}\gtrsim 10^{23}~\Ncc$), the break strength decreases toward 
$n_{\rm H}\simeq 10^{11}~\cc$ and rises again at higher densities.
This non-monotonic trend arises from the complex interplay of physical processes that govern the population of 
hydrogen atoms in the excited states.

Figure.~\ref{fig:SED_BD_nH} illustrates how the Balmer decrement varies with the gas density.
The left panel presents the emitted continuum and line spectra calculated without dust grains.
Here, the column density is fixed at $N_{\rm H}=10^{24}~\Ncc$, while the volume density varies over $\log(n_{\rm H}/\cc)=2$, 
3, 5, 7, 9, and 11 from the bottom to the top. 
The wavelength range of $4000~{\rm \AA} \leq \lambda \leq 6800~{\rm \AA}$ includes 
Balmer lines ($\Hg$, $\Hb$, and $\Ha$) and [\ion{O}{3}] emission, marked by vertical dashed lines.
The numbers beside each spectrum indicate the Balmer decrement, $F({\rm H\alpha})/F({\rm H\beta})$,
measured directly from the model SEDs.
At low densities ($n_{\rm H}\simeq 10^2~\cc$), the Balmer decrement is consistent with the Case~B value of $2.86$.
As density increases, the Balmer decrement first declines slightly, and then rises greatly to a maximum of
$97.74$ at $n_{\rm H}=10^7~\cc$. 
Beyond this peak, it decreases again toward higher densities ($n_{\rm H}=10^{8-11}~\cc$). 
These examples demonstrate that while the Balmer decrement reproduces the Case~B value in the low density limit,
it can deviate by orders of magnitude at higher densities even in the absence of dust extinction.

The right panel summarizes this trend by showing the Balmer decrement as a function of gas density
for $N_{\rm H}=10^{22}-10^{24}~\Ncc$, along with the Case~B value indicated by a horizontal dashed line.
The overall behavior reflects changes in the optical depths of H$\alpha$, H$\beta$, and Pa$\alpha$,
which we discuss in Section~\ref{subsec:Resonance scatting}.

\subsection{Physical mechanisms for modifying Balmer decrements} 
\label{subsec:Resonance scatting}

\begin{figure}[t]
    \includegraphics[width=85mm]{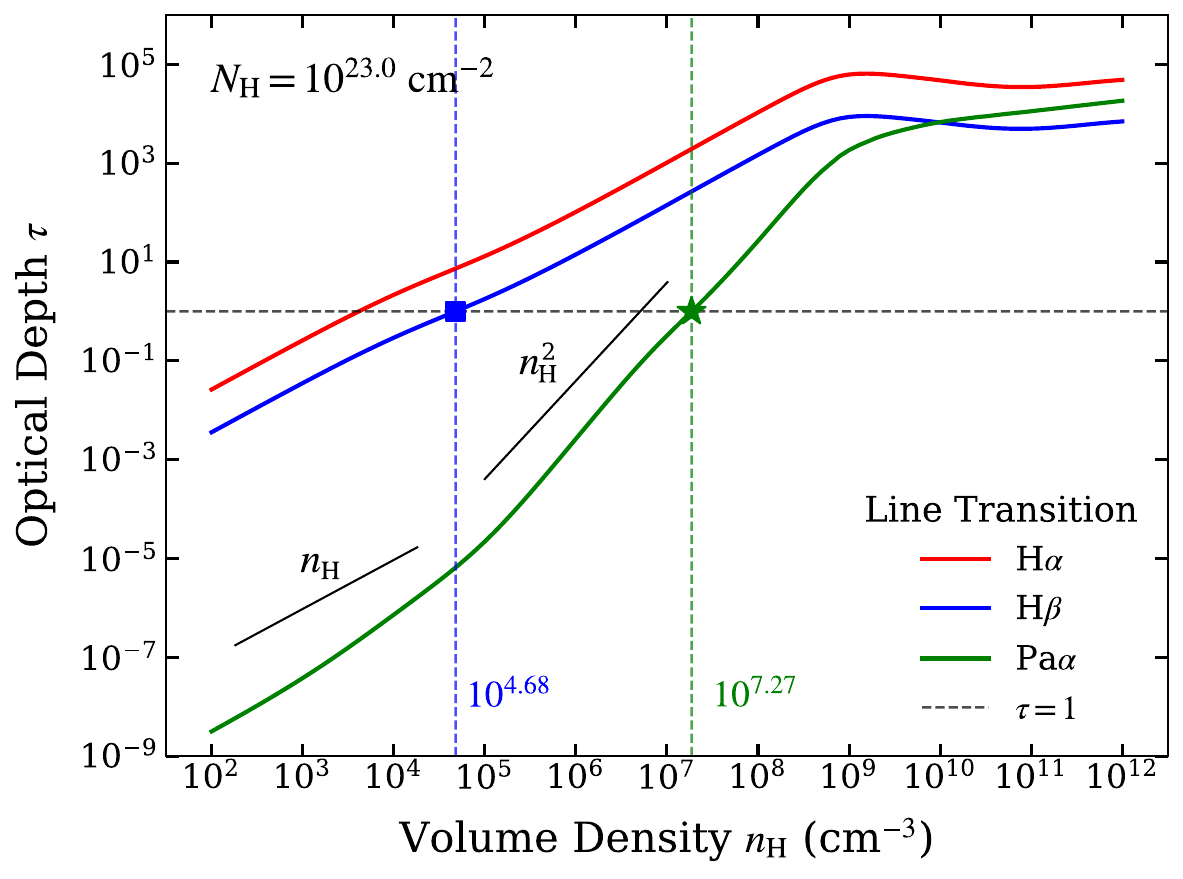}
\vspace{-4mm}
\caption{
Optical depths of the hydrogen recombination lines (H$\alpha$, H$\beta$, and Pa$\alpha$) 
as a function of hydrogen volume density $n_{\mathrm{H}}$ for a fixed hydrogen column density 
$N_{\mathrm{H}} = 10^{23}\,\mathrm{cm^{-2}}$.
With a fixed $\NH$, the optical depths of Balmer lines ($\tau_{\Ha}$ and $\tau_{\Hb}$) increase with 
$\propto \nH$ due to collisional excitation to the $n=2$ level.
While $\tau_{\Paa}$ follows a similar trend in the low density regime, the value rapidly increases as 
$\propto n_{\rm H}^2$ when the gas slab becomes opaque to $\Hb$ line and its resonance scattering operates (see text).
At high densities above $n_{\rm H}\simeq 10^9~\cc$, all the optical depths approach their LTE values. 
Square and star symbols denote the densities at which $\tau_{\Hb}=1$ and $\tau_{\mathrm{Pa}\alpha}=1$, respectively, 
with the corresponding $\nH$ values.
The relative ordering of the line optical depths explains the Balmer decrement trends presented in the right panel of Figure~\ref{fig:SED_BD_nH}.
}
\vspace{3mm}
\label{fig:tau_nH}
\end{figure}

Next, we describe the physical processes that determine hydrogen level populations in dense gas.
To examine hydrogen excitation, we first show in Figure~\ref{fig:tau_nH}, the optical depths of the hydrogen recombination lines 
H$\alpha$, H$\beta$, and Pa$\alpha$ 
as a function of hydrogen density for a fixed column density of $N_{\rm H}=10^{23}~\Ncc$. 
The optical depth of a transition from an upper level $u$ to a lower level $l$ is given by
\begin{equation}
\tau_{ul} = \frac{A_{ul}\lambda_{ul}^3}{8\pi} \cdot \frac{n_l}{n_{\rm H}}\left(\frac{g_u}{g_l} - \frac{n_u}{n_l}\right) \frac{N_{\rm H}}{\sigma_V},
\end{equation}
where $A_{ul}$ is the spontaneous radiative decay rate,
$\lambda_{ul}$ ($\nu_{ul}$) is the wavelength (frequency) of the emitted line, $n_{u(l)}$ is the population density of the upper (lower) energy level, 
$g_{u(l)}$ is the corresponding statistical weight, $\sigma_V=\sqrt{2k_{\rm B}T/m_{\rm H}}$ is 
the thermal velocity of hydrogen at a temperature of $T$, $k_{\rm B}$ is the Boltzmann constant, and $m_{\rm H}$ is the hydrogen mass.

In the low density regime, the population of the $n=2$ level is set by the balance between collisional excitation from 
the ground state ($n=1$) and two-photon decay from the $2s$ state under the Case~B conditions, where the Ly$\alpha$ transition ($2p\rightarrow 1s$)
is effectively suppressed due to enormous numbers of scattering and the $2p$ state is collisionally converted to the $2s$ state before Ly$\alpha$ escaping
from the cloud \citep{Spitzer&Greenstein1951,Omukai2001,Dijkstra_2016}.
As a result, collisional excitation enhances the excited-level populations, yielding $n_2 \propto n_{\rm H}^2$,
and thus $\tau_{\rm H\alpha(H\beta)} \propto n_{\rm H}$.
Similarly, the $n=3$ level is collisionally excited from the ground state, so that $\tau_{\rm Pa\alpha} \propto n_{\rm H}$ 
as shown in Figure~\ref{fig:tau_nH}.

As the gas density exceeds $n_{\rm H}\simeq 5\times 10^4~\cc$, the H$\beta$ line becomes optically thick (blue square and vertical line),
as also indicated in Figure~\ref{fig:SED_BD_nH}.
In this regime, photons near the H$\beta$ transition energy are repeatedly absorbed and re-emitted by hydrogen atoms in the $n=2$ level, 
exciting them to the $n=4$ level and back.
During multiple resonance scattering, some electrons in the $n=4$ level may follow an alternative decay path:
instead of directly returning to $n=2$ (emitting an H$\beta$ photon), they can first decay to $n=3$ and then from $n=3$ to $n=2$ 
(emitting a Pa$\alpha$ and an H$\alpha$ photon).
This cascade effectively converts a fraction of the H$\beta$ emission into H$\alpha$, thereby enhancing the Balmer decrement 
\citep{Netzer1975,Krolik&McKee1978,Chang2025}.
Since this pathway also increases the population of the $n=3$ level, the $n_3/n_1$ ratio rises non-linearly ($\propto n_{\rm H}^2$), 
and the optical depth of Pa$\alpha$ grows accordingly (see Figure~\ref{fig:tau_nH}).
As the density further increases, the Pa$\alpha$ optical depth exceeds unity at $n_{\rm H}\simeq 2\times 10^7~\cc$ (green star and vertical line),
where the conversion of $\Hb \rightarrow \Paa \rightarrow \Ha$ becomes inefficient because $\Paa$ photons are trapped.
As a result, the Balmer decrement decreases again, marking the end of the enhanced Balmer decrement phase.

Following the qualitative explanation above, we now provide a quantitative evaluation of the Balmer decrement in the regimes where 
the $F(\Ha)/F(\Hb)$ ratio rapidly rises and subsequently declines at $10^5\lesssim \nH/\cc \lesssim 10^{10}$.
The ratio is expressed as
\begin{equation}
    \frac{F(\Ha)}{F(\Hb)}=\frac{\lambda_{42}\bar{A}_{32}n_3\beta_{32}}{\lambda_{32}\bar{A}_{42}n_4\beta_{42}}
\end{equation}
where $\beta_{ul}$ denotes the photon escape probability approximated as 
\begin{equation}
\beta_{ul}
\simeq 
\dfrac{1-\exp(-\tau_{ul})}{\tau_{ul}} \approx 0.1\left(\frac{\tau_{ul}}{10}\right)^{-1},
\end{equation}
\citep{Mihalas1978,deJong1980}.
The effective spontaneous decay rate $\bar{A}_{ul}$ is computed by averaging over the sub-levels with 
orbital quantum number ($s$, $p$, $d,\dots$) 
as $\bar{A}_{ul}=\Sigma_{i} A_{il}n_i/\Sigma_i n_i$, where $i$ runs over all sub-levels of the principal quantum number $n=u$
(i.e., $n_u=\Sigma_i n_i$).
As the $\Hb$ line becomes optically thick and undergoes resonance scattering, the boosted $\Ha$ emission is led by
the $\Paa$ transition. Thus, one can approximate that the $n=3$ level population is determined by
\begin{equation}
    \bar{A}_{32}n_3\beta_{32}=\bar{A}_{43}n_4\beta_{43}.
\end{equation}
Therefore, we obtain
\begin{equation}
\frac{F(\Ha)}{F(\Hb)}=\frac{\lambda_{42}}{\lambda_{32}}\frac{\bar{A}_{43}n_4\beta_{43}}
{\bar{A}_{42}n_4\beta_{42}}= \mathscr{D}\beta_{43}\beta_{42}^{-1},
\end{equation}
where the numerical coefficient is computed as $\mathscr{D}=0.54$ using the CLOUDY output of the level populations 
in the $4s$, $4p$, $4d$, and $4f$ states for $\nH =10^{6}~\cc$ and $\NH=10^{23}~\Ncc$.
In this parameter range, the $4p$ sub-level dominates the total $n=4$ population, so that $\Paa$ primarily arises from 
the $4p\rightarrow 3s/3d$ transitions, while $\Ha$ originates from the $3s/3d\rightarrow 2p$ transitions. 
We note that with only the contribution from the $4p$ state, the numerical coefficient is $\mathscr{D}=0.46$.

In summary, the Balmer decrement can be approximated as 
\begin{equation}
\frac{F(\Ha)}{F(\Hb)}
\simeq 
\left\{
\begin{array}{ll}
5.4\left(\dfrac{\tau_{\rm H\beta}}{10}\right) & ~{\rm for}~\tau_{\rm Pa\alpha}<1,\\[8pt]
54~\left(\dfrac{\tau_{\rm H\beta}}{10^3}\right)\left(\dfrac{\tau_{\rm Pa\alpha}}{10}\right)^{-1} &~{\rm for}~\tau_{\rm Pa\alpha}\gg 1.
\end{array}
\right.
\end{equation}
Since the optical depths scale with $\nH$ and $\NH(>10^{22}~\Ncc)$ as shown in Figure~\ref{fig:tau_nH},
\begin{align}
 \tau_{\rm H\beta}\simeq 13.3~\left(\frac{n_{\rm H}}{10^6~\cc}\right)\left(\frac{\NH}{10^{23}~\Ncc}\right)^{0.53},\\[5pt]
 \tau_{\rm Pa\alpha}\simeq 22~\left(\frac{n_{\rm H}}{10^8~\cc}\right)^2\left(\frac{\NH}{10^{23}~\Ncc}\right)^{0.66},
\end{align}
within the gas slab where $\Hb$ resonance scattering frequently occurs,
the above expressions reduce to
\begin{equation}
\frac{F(\Ha)}{F(\Hb)}
\simeq 
\left\{
\begin{array}{ll}
7.2~\left(\dfrac{n_{\rm H}}{10^6~\cc}\right) \left(\dfrac{\NH}{10^{23}~\Ncc}\right)^{0.53} \\[5pt]
~~~~~~~~~~~~~~~~~~~~~~~~~~~~~~~~{\rm for}~\tau_{\rm Pa\alpha}<1,\\[8pt]
33~\left(\dfrac{n_{\rm H}}{10^8~\cc}\right)^{-1} \left(\dfrac{\NH}{10^{23}~\Ncc}\right)^{-0.13} \\[5pt]
~~~~~~~~~~~~~~~~~~~~~~~~~~~~~~~~{\rm for}~\tau_{\rm Pa\alpha}\gg 1,
\end{array}
\right.
\label{eq:BD_LTE_nN}
\end{equation}
The analytical formula of the Balmer decrement agrees well with the numerical results shown in Figure~\ref{fig:SED_BD_nH},
while the high-density part underestimates the flux ratio by a factor of 1.8.
The column-density dependence is also broadly consistent with the numerical results.

\subsection{Correlation between Balmer break and Balmer decrement\label{subsec:correlation between BB and BD}}

\begin{figure}[t]
    \centering
    \includegraphics[width=85mm]{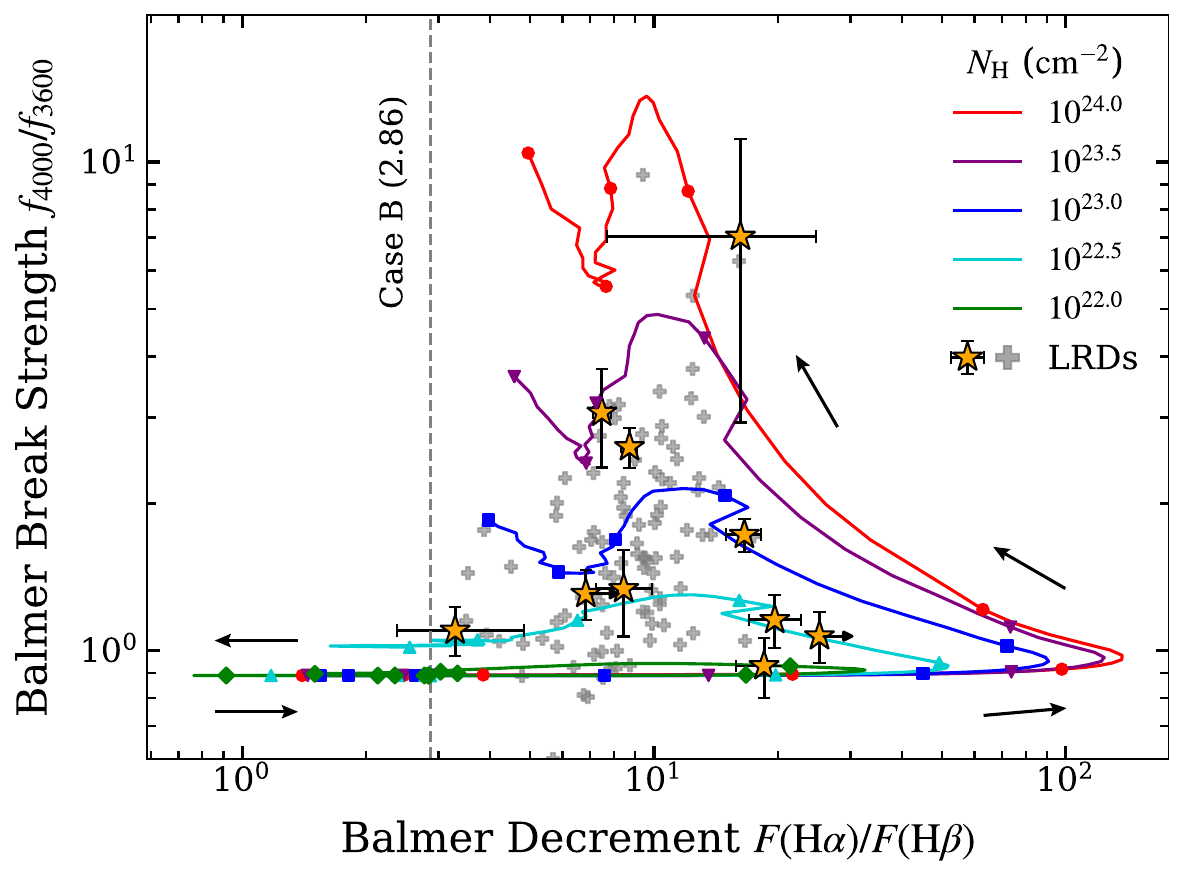}
    \caption{
        Correlations between the Balmer break strength and the Balmer decrement for different hydrogen column densities 
        of $N_{\mathrm{H}} = 10^{22}-10^{24}~\Ncc$.
        Each curve traces increasing hydrogen volume density $n_{\mathrm{H}}$ along the sequence. The vertical line marks the canonical Case~B recombination value for reference. LRDs with measured Balmer break and Balmer decrement are overlaid to demonstrate the diagnostic capability of this diagram for constraining both the volume density and column density of high density clouds. Orange star symbols represent LRDs for which the broad and narrow components have been spectroscopically decomposed, including objects discussed in \cite{Nikopoulos2025} and several well-known LRDs (see Table~\ref{tab:BB_BD}).
        For these sources, the Balmer decrements of the broad-line component are shown. 
        Gray cross symbols denote an LRD sample with Balmer decrements measured from the total emission line fluxes \citep{deGraaff25b}.
    }
    \label{fig:BB_BD}
\end{figure}

Figure~\ref{fig:BB_BD} illustrates the correlation between the Balmer break strength 
and the Balmer decrement for a grid of photoionization models spanning a wide range of hydrogen volume densities and column densities.
Each curve corresponds to a fixed $\NH$ and filled symbols mark integer steps of $\Delta \log \nH=1$ along each curve.

At low densities ($n_{\rm H} \lesssim 10^{7}~\cc$), the Balmer break is negligible for all column densities, 
yielding nearly horizontal tracks. 
In the low-density limit ($n_{\rm H} \simeq 10^{2-4}~\cc$), hydrogen level populations are dominated by radiative cascades,
and thus the Balmer decrement remains close to the Case~B value of $\Ha /\Hb \simeq 2.86$. 
As the density increases, the gas first becomes opaque to $\Ha$ line and the Balmer decrement decreases. 
Thus, for each $\NH$, the tracks start at the Case~B point and move leftward until $\Hb$ also becomes optically thick at 
$n_{\rm H}\sim 10^{5}~\cc$, where the decrement reaches its minimum ($\Ha/\Hb\simeq0.8$).

Once $\Hb$ becomes optically thick, resonance scattering (Section~\ref{subsec:Resonance scatting}) elevates the Balmer decrement,
causing each sequence to bend rightward.
This increase continues until the $\Paa$ transition becomes optically thick at $\nH\simeq 10^7~\cc$, where
the boosted Balmer decrement terminates.
At the same time, under Case~B conditions, where Lyman transitions are effectively suppressed, the $n=2$ population is maintained through 
collisional excitation and approximated as $n_{2}\propto n_{\rm H}^2$ via balance with two-photon decay.
At intermediate gas densities ($n_{\rm H} \sim 10^{8-9}~\cc$), the number of atomic hydrogen at $n=2$ becomes high enough to generate a measurable Balmer break.
The combined evolution of the Balmer break and the Balmer decrement drives the tracks upward and to the left.

At higher densities ($n_{\rm H} \gtrsim 10^{9}~\cc$), the hydrogen level populations begin to approach their LTE values. 
In this regime, the Balmer decrement converges toward a nearly constant value that is insensitive to the column density choice 
once $\NH\gtrsim 10^{23}~\Ncc$ (see the analytical form in Equation~\ref{eq:BD_LTE_nN}). 
In contrast, the Balmer break strength retains a dependence on $\NH$, causing the model tracks to separate at high densities.
This response breaks the degeneracy present at $n_{\rm H}\lesssim 10^8~\cc$ and provides a two-dimensional diagnostic of both the hydrogen volume density 
and column density when the Balmer break and Balmer decrement are jointly measured.

In Figure~\ref{fig:BB_BD}, LRD samples with measured Balmer break and decrement values are overlaid. 
The orange star symbols represent LRDs for which the broad and narrow components have been spectroscopically decomposed, including objects discussed in \cite{Nikopoulos2025} and several well-known LRDs. The observational data for these sources are listed in Table~\ref{tab:BB_BD}. The gray cross symbols denote the LRD sample for which the Balmer decrements are measured from the total emission line fluxes \citep{deGraaff25b}. 
We note that the Balmer decrement values for the broad $\Ha$ and $\Hb$ emission of the gray symbols are likely higher if the line decomposition is performed (see \citealt{Nikopoulos2025}).
Both samples occupy regions corresponding to well-defined combinations to $\nH$ and $\NH$, demonstrating the practical diagnostic power of this method.
Moreover, the absence of models and observational data in the upper-right region of Figure~\ref{fig:BB_BD} suggests a physically 
inaccessible zone where neither radiative nor collisional equilibrium can be maintained.

\section{Discussion} \label{sec:discussion}

\subsection{Balmer decrement: an estimator for dust?\label{subsec:Balmer decrement as an estimator for dust}}

\begin{figure}[t]
    \centering
    \includegraphics[width=86mm]{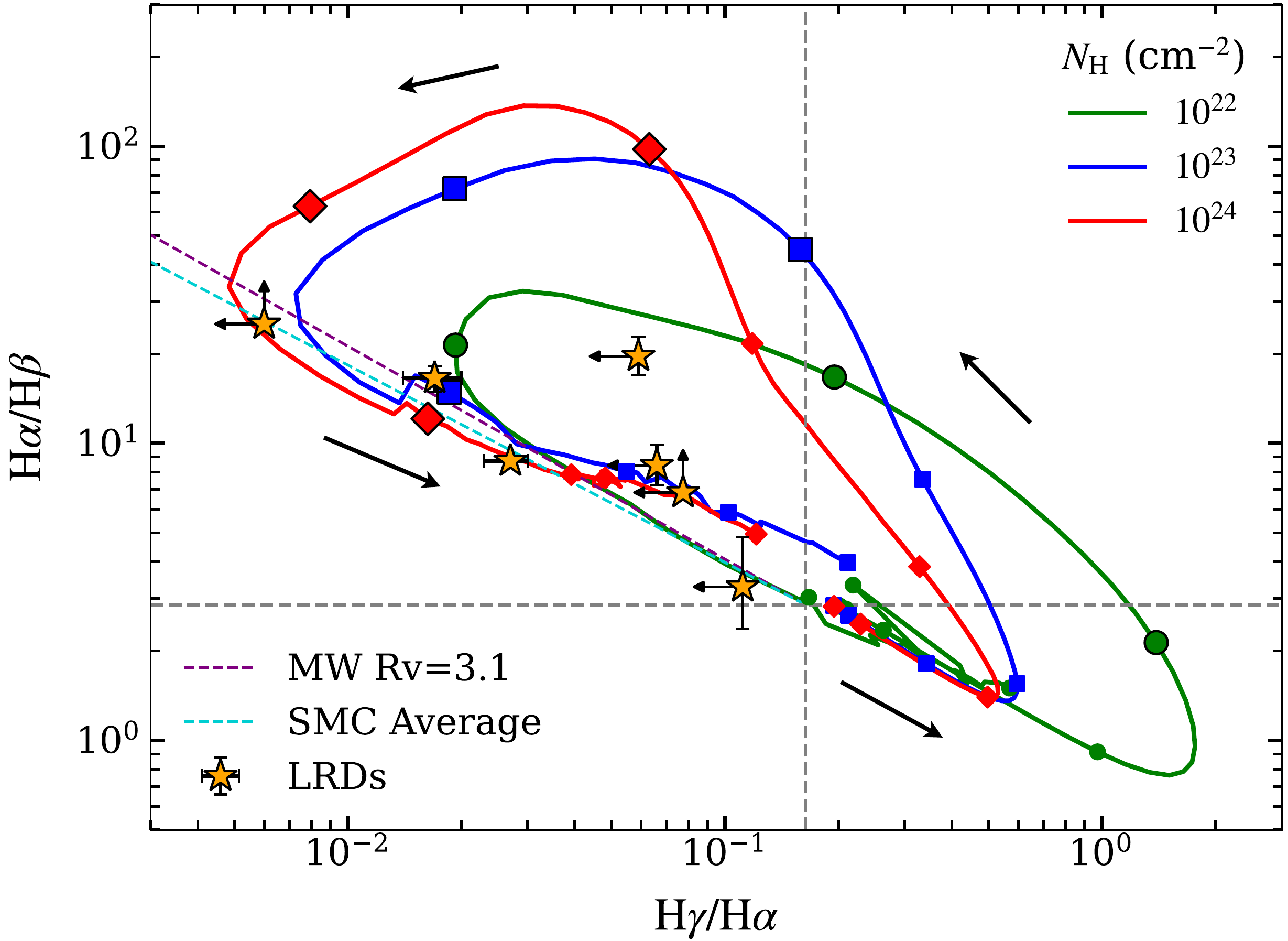}
    \caption{
        Balmer line ratio diagram showing $\Ha/\Hb$ versus $\Hg/\Ha$ for three different hydrogen column densities ($\NH= 10^{22}$, $10^{23}$, and $10^{24}~\Ncc$). Each curve traces increasing hydrogen volume density from $\nH=10^{2}$ to $10^{12}\,\cc$ along a counterclockwise sequence, beginning at the Case~B locus indicated by the intersection of the two dashed gray lines. Larger markers highlight densities of $\nH=10^7$, $10^8$, and $10^9~\cc$. Star symbols represent the observed broad line ratios compiled by \cite{Nikopoulos2025}. Two representative dust extinction curves are plotted for comparison. The ratios observed in these LRDs are naturally reproduced by models with 
        $\nH = 10^{9}-10^{12}~\cc$.
    }
    \label{fig:HaHb_vs_HgHa}
\end{figure}

In previous studies, deviations of the Balmer decrement from the canonical Case~B values have generally 
been interpreted as a result of dust reddening. 
Figure~\ref{fig:HaHb_vs_HgHa} presents the broad Balmer-line flux ratios $\Ha$/$\Hb$ and $\Hg$/$\Ha$. 
The orange star symbols represent the observed LRD sample compiled by \cite{Nikopoulos2025},
including error bars and upper limits.
These measurements are significantly deviated from the Case~B values (horizontal and vertical dashed lines)
and occupy the upper-left region of the diagram.
Such offsets are consistent with dust-reddening interpretations and can indeed be reproduced by representative extinction curves 
(the Milky Way curve with $R_V=3.1$, and the SMC average curve), provided that moderate to substantial dust obscuration 
($A_V\simeq1-8 ~{\rm mag}$) is present in the BLR \citep{Nikopoulos2025}.
However, this level of dust obscuration inevitably reprocesses a large fraction of the absorbed radiation energy 
into the NIR band \citep{Z.Li_2025,chen2025dustbudgetcrisislittle}, which is strongly constrained by JWST MIRI observations 
\citep{Williams24,Akins_2025,Setton25,Casey2025}.

In contrast, dense line-emitting clouds naturally produce Balmer decrements higher than the Case~B values without dust.
Figure~\ref{fig:HaHb_vs_HgHa} presents the evolutionary tracks of Balmer-line ratios for three different column densities 
($\NH= 10^{22}$, $10^{23}$, and $10^{24}~\Ncc$). 
For each $\NH$, the curves trace increasing hydrogen volume density from $\nH = 10^2$ to $10^{12}~\cc$ 
in a counter-clockwise direction, beginning from the Case~B intersection of the two dashed lines. 
Filled symbols indicate steps of $\Delta \log \nH=1$, with larger markers highlighting $\nH=10^7$, $10^8$, and $10^9~\cc$
(lower- and higher-density cases are clustered too closely to distinguish clearly).
A similar rotating behavior has been already found in a pioneer study by \citet{Netzer1975},
where the optical depths of Ly$\alpha$ and $\Ha$ vary under a constant electron volume density.

Intriguingly, at $n_{\rm H}\gtrsim 10^{9}~\cc$, the model predictions converge toward the dust-reddening sequences without invoking dust.
This behavior closely resembles the broad-line Balmer decrements observed in LRDs \citep{Nikopoulos2025}, suggesting that 
high-density gas alone can account for the measurements.
This apparent convergence trend is understood in the following reason.
In the LTE regime, the line flux ratios approach an asymptotic value of
$F(\Ha)/F(\Hb)=2.90$ and $F(\Ha)/F(\Hg)=5.75$ at $T=6000~\K$, respectively, as discussed in Appendix~\ref{sec:appB}.
Both values are quite consistent with the Case~B values of $2.86$ and $6.10$ \citep{Osterbrock&Ferland2006}.
In the analytical calculation, the temperature dependence of the flux ratio is given in Equation~(\ref{eq:HaHg_LTE}) as
\begin{align}
\frac{F(\Ha)}{F({\rm H}n)}\propto \exp\left(\frac{h\nu_{3n}}{k_{\rm B}T}\right), 
\end{align}
Eliminating the temperature $T$, one can derive a single power-law relation of 
\begin{align}
\frac{F(\Ha)}{F(\Hb)} \propto \left[\frac{F(\Hg)}{F(\Ha)}\right]^{-\nu_{34}/\nu_{35}},
\label{eq:LTE_dust}
\end{align}
where $\nu_{34}/\nu_{35}=0.684$.
The slope appears close to those of the dust extinction curves (the dashed lines in Figure~\ref{fig:HaHb_vs_HgHa}).
Taking higher-order Balmer lines, the slope approaches unity.

We conclude that the Balmer line ratios are not a reliable tracer of dust extinction in the BLRs of LRDs. 
Their deviations from the Case~B values in multiple line ratios can be naturally explained by 
radiative transfer effects operating in dense gaseous media. 
Thus, dust estimates based only on Balmer decrements can be incorrect for BLR gas.
Within the dense-gas scenario, the combination of the Balmer break strength and multiple Balmer-line ratios 
(e.g., $\Ha/\Hb$ versus $\Hg/\Ha$) provides a powerful diagnostic for constraining the physical conditions of the BLRs in LRDs.

\subsection{High EWs of broad H$\alpha$ emission}

\begin{figure}[t]
    \centering
    \includegraphics[width=86mm]{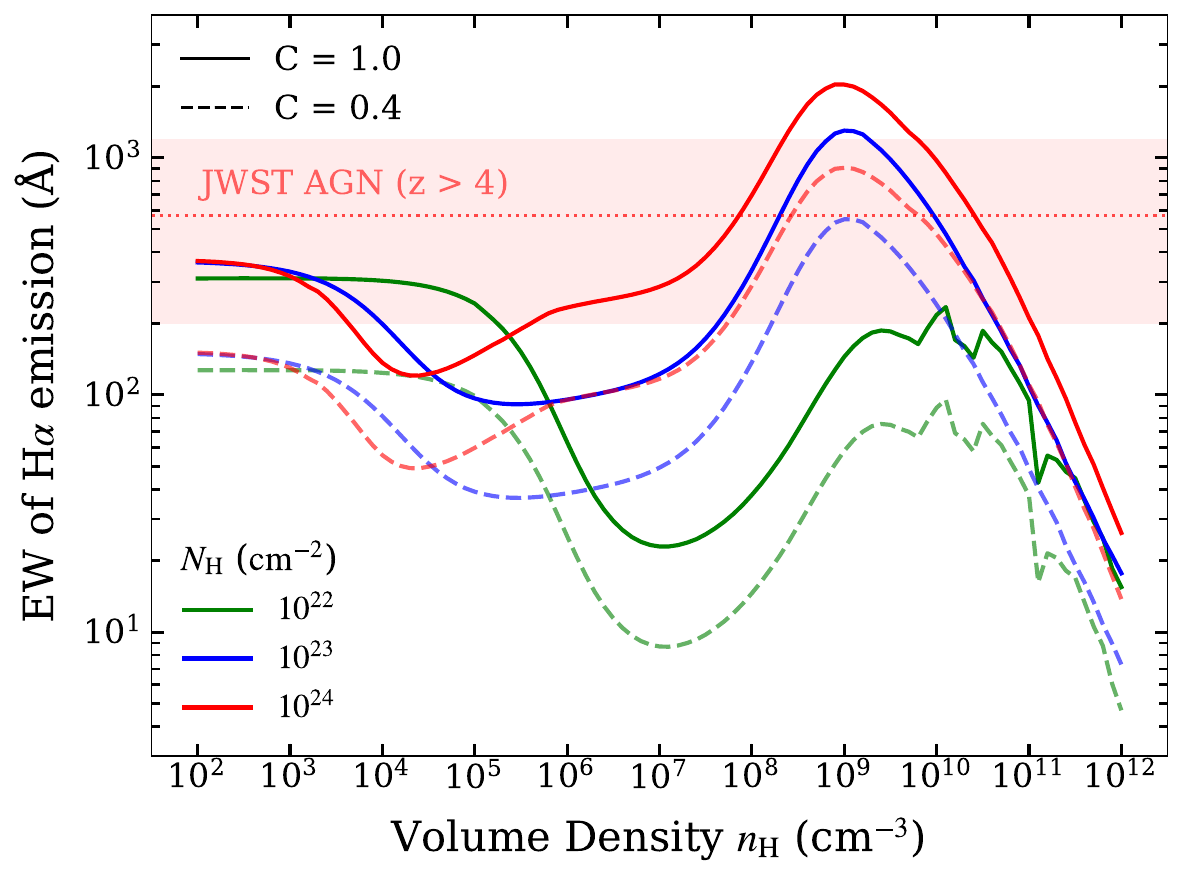}
    \caption{
        Rest-frame EW of $\Ha$ emission as a function of hydrogen volume density for three column densities $\NH=10^{22}-10^{24}\,\Ncc$. Solid and dashed curves correspond to covering factions $C=1.0$ and $C=0.4$, respectively. Red dotted line and shaded region show the observed median and distribution of broad $\Ha$ EW for high-redshift AGNs from JWST \citep{Maiolino2025}. The extremely high EW of broad $\Ha$ emission can only be produced by covering faction $C\sim1$ and high density gas environment.
    }
    \label{fig:EW}
\end{figure}

JWST observations reveal that LRDs exhibit not only unusual Balmer break strengths and Balmer decrements, both defined by flux ratios, 
but also abnormally strong $\Ha$ emission.
The median (rest-frame) EW of the broad $\Ha$ line for JWST-identified, high-redshift AGN (${\rm EW}_{\Ha}=570~{\rm \AA}$; \citealt{Maiolino2025}) is 
$\sim 3$ times larger than that of low-redshift AGNs (${\rm EW}_{\Ha}=200~{\rm \AA}$; \citealt{Lusso_2020}). 
Because the EW of a recombination line is highly sensitive to the covering factor 
of the line-emitting gas when the incident SED shape of the ionizing radiation is fixed, larger EWs imply a high covering fraction of clouds, 
substantially higher than $C \sim 10-50\%$ typically inferred for BLRs in low-redshift, optically/UV-selected AGNs 
\citep{Netzer1990,Dunn2007,Baskin2018}.
This naturally points to a denser, more heavily enshrouded environment during the early phases of SMBH growth 
\citep[e.g.,][]{Inayoshi&Maiolino2025}.

Figure~\ref{fig:EW} illustrates the dependence of the broad $\Ha$ EW on the hydrogen volume density and column density.
The red dotted line and the shade region show the observed median and distribution of the broad $\Ha$ EWs for 
high-redshift AGNs including LRDs \citep{Maiolino2025}.
In comparison to the model EWs, the largest observed values cannot be explained by a high covering faction alone. 
Even with $C=1$, the predicted EWs remain too small unless the gas resides in a dense environment with $\nH\simeq10^{8-10}~\cc$ 
and $\NH\simeq10^{23-24}~\Ncc$.
Therefore, both a covering fraction approaching unity and high-density BLR gas are required to reproduce 
the upper envelope of the observed EW distribution.

\subsection{Mass, energy, and metals in BLR clouds}

The characteristic Balmer-transition features observed in many LRDs suggest that they are 
powered by AGNs enshrouded by dense gaseous environments.
We here briefly examine the corresponding mass and energy budget, based on the volume and column densities 
inferred from the elevated Balmer decrements, Balmer break strengths, and Balmer broad-line EWs.

We assume that BLR clouds in an LRD are distributed in a spherical shell with a width of $\Delta R$ at 
a distance $R_{\rm BLR}$ from the central BH.
Since the clouds are dense and opaque to Balmer lines (and partially to $\Paa$), 
the required physical thickness of each cloud is 
\begin{equation}
    \Delta R \simeq \frac{\NH}{\nH}\simeq10^{14}~{\rm cm} ~ N_{\rm H, 23} n_{\rm H, 9}^{-1},
\end{equation}
where $n_{\rm H, 9} =\nH/(10^9 ~\cc)$ and $N_{\rm H, 23}=\NH/(10^{23}~\Ncc)$. 
The characteristic BLR distance is assumed to follow the $R-L$ relation calibrated from reverberation mapping
\citep[e.g.,][]{Greene&Ho2005,Kaspi2000,Peterson2004}:
\begin{align}
    R_{\rm BLR}
    \simeq 0.025~\pc~L_{44}^{0.64},
\end{align}
where $L_{44}=L_{\rm AGN}/(10^{44}~{\rm erg~s^{-1}})$ and $L_{\rm AGN}$ is the rest-frame $5100~{\rm \AA}$ continuum luminosity.
The total gas mass contained in the BLR shell is then
\begin{align}
    M_{\rm BLR} &= 4\pi CR^2 \NH m_{\rm H}\simeq 
    6.3 ~ \msun ~C L_{44}^{1.28}N_{\rm H, 23}.
\end{align}
This mass estimate includes ionized, partially ionized, and neutral components within each cloud.
Compared with typical BLR masses of local AGNs ($\sim 10^{3-4}~\msun$; \citealt{Baldwin2003}), 
the inferred BLR mass for LRDs is substantially lower because of their lower luminosities. 
For luminous quasars with $L_{5100}\sim 10^{47}~{\rm erg~s}^{-1}$, as considered in \citet{Baldwin2003},
the BLR mass would increase to $\sim \mathcal{O}(10^4~\msun)$.

%
Assuming that the cloud thickness $\Delta R$ represents the diameter of its cross section, 
the mass of a single cloud becomes
\begin{equation}
    M_{\rm cl}=\frac{4\pi}{3}\left(\frac{\Delta R}{2}\right)^3\nH m_{\rm H}\simeq4.4\times10^{-7}~M_\odot N_{\rm H, 23}^3 n_{\rm H, 9}^{-2}.
\end{equation}
The total number of clouds required to supply the BLR mass is therefore
\begin{equation}
    N_{\rm cl}=\frac{M_{\rm BLR}}{M_{\rm cl}}\simeq
    1.4\times10^7~C L_{44}^{1.28} N_{\rm H, 23}^{-2} n_{\rm H, 9}^{2},
\end{equation}
This estimate is consistent with the empirical lower limit of $\gtrsim 3\times 10^6$ clouds 
inferred from high-$S/N$, high resolution spectroscopy \citep{Arav1997}.
Future high-$S/N$ JWST spectra of LRDs may further constrain the minimum cloud number \citep[e.g.,][]{Rusakov2025}.

Such low masses of BLR clouds inferred for LRDs (and for faint AGNs that show similar Balmer-transition signatures)
also provides insight into their metal enrichment.
LRDs exhibit weak or absent metal-line emission such as \ion{C}{4}, \ion{C}{3}], and \ion{Mg}{2}, implying either 
low metallicity or a soft ionizing continuum \citep{Labbe2024_metal,Wang2025_misshardphoton}.
For the fiducial metallicity of $Z=0.1Z_\odot$ adopted in our CLOUDY calculations, the total metal mass contained in the BLR gas is
\begin{align}
    M_{\rm metal} 
    \simeq 1.3 \times 10^{-2}~\msun  CN_{\rm H, 23} L_{44}^{1.28} \left(\frac{Z}{0.1 Z_\odot}\right).
\end{align}
However, a single supernova typically ejects $\sim 0.1-1~\msun$ of metals into its surroundinging medium \citep{Nomoto2006}. 
Thus, even one recent SN within the BLR would overproduce metals relative to the small mass inferred above.
The metallicity implied by the Balmer-transition modeling therefore places a stringent upper limit on the supernova rate
and more generally, the recent star-formation activity at a vicinity of the BLR \citep{Inayoshi_2025S}.

\subsection{Application to narrow-line emission}

In addition to using the Balmer decrement to diagnose the BLR properties of LRDs, 
a similar analysis can be extended to the narrow line regions (NLRs). 
NLR clouds generally have lower column and volume densities compared to those in BLRs.
As shown in the right panel of Figure~\ref{fig:SED_BD_nH}, clouds with 
$N_{\rm H}\simeq 10^{22}~\Ncc$ and $n_{\rm H}\simeq 10^{5-7}~\cc$ can produce 
Balmer decrements below the canonical Case~B value ($\Ha/\Hb<3$), which is a phenomenon known as the Balmer decrement anomaly. 
This behavior has already been reported in several observations of star-forming galaxies \citep[e.g.,][]{Cameron24,Topping24}.

\cite{Yanagisawa24_BalmerAnomaly} showed that similarly low $\Ha/\Hb$ ratios can be raised
even in clouds with $n_{\rm H}=10^2-10^5~\cc$ and relatively low column densities.
They discussed two possible mechanisms.
In the first scenario, the optical depths of the Lyman-series transitions are lower than assumed in Case B.
When the cloud is optically thick only in Ly$\alpha$-Ly$\gamma$, but optically thin in higher Lyman transitions,
Ly$\gamma$ photons can be converted efficiently into H$\beta$ while H$\alpha$ remains unaffected, thereby reducing the Balmer decrement.
This effect gradually disappears at higher column densities once the cloud enters the Case B regime,
where all Lyman lines become optically thick.
In the second scenario, an ionized nebula surrounded by a layer of excited neutral hydrogen. 
The surrounding neutral gas absorbs $\Ha$ more effectively than $\Hb$, leading to a $\Ha/\Hb$ value below the Case~B value.
In contrast to these low-density mechanisms, our analysis explores a higher-density regime at $n_{\rm H}\simeq 10^5-10^7~\cc$, 
where the Balmer decrement is reduced to $\Ha/\Hb \lesssim 2$ due to the optical thickness of $\Ha$, 
significantly lower than the minimum values identified by earlier work (but see \citealt{Netzer1975}).

Given that the gas densities required to yield Balmer decrement anomalies are comparable to the critical density of low-ionization forbidden lines 
(e.g., $n_{[\rm OIII]}\simeq 7\times 10^5~\cc$ and $n_{[\rm NII]}\simeq 10^4~\cc$), we expect these forbidden transitions to be collisionally suppressed in systems with $\Ha/\Hb<2$.
A systematic examination of the correlation between the Balmer decrement, the Balmer break strength, and the weakness of forbidden lines will therefore provide a more comprehensive diagnostic of the structure of NLRs in high-redshift AGNs and LRDs. 
This will be left for future work.

\section{Summary} \label{sec:summary}

The newly discovered AGN population known as LRDs exhibit distinctive Balmer-transition signatures,
including large Balmer decrements, prominent Balmer absorption on top of broad-emission lines, pronounced Balmer breaks, 
and large (rest-frame) EWs of broad $\Ha$ emission.
All the characteristics suggest that LRDs are rapidly growing BHs enshrouded by dense gas.
Using CLOUDY-based radiative transfer calculations through dust-free, high density gas, we have investigated 
the physical origin of these Balmer features. Our main conclusions are as follows:

\begin{itemize}
\item Large departures of the $\Ha/\Hb$ line flux ratio from the Case~B value observed in LRDs (and some AGNs) do not 
necessarily require dust reddening.
Radiative transfer effects in dense gas, including high optical depths and resonance scattering of Balmer lines, 
can naturally generate Balmer decrements far above the canonical Case~B prediction, reaching maximum values of 
$20-100$ at $\nH\simeq 10^7-10^9~\cc$ and $\NH \gtrsim 3\times  10^{22}~\Ncc$ (see Figure~\ref{fig:SED_BD_nH} and 
Equation~\ref{eq:BD_LTE_nN}).

\item At high densities ($\nH \gtrsim 10^{7}-10^{9}~\cc$), elevated multiple Balmer-line ratios 
(e.g., $\Ha/\Hb$, $\Hg/\Ha$) converge to values that mimic dust reddening (see Figure~\ref{fig:HaHb_vs_HgHa} and Equation~\ref{eq:LTE_dust}).
The broad-line Balmer decrements observed in LRDs therefore more plausibly trace dense BLR gas rather than heavy dust extinction.
This conclusion is further supported by the weak infrared dust emission seen in JWST/MIRI observations.

\item If the Balmer break and broad Balmer lines on LRD spectra originate from the same dense medium, their strengths are physically linked.
This connection provides a practical diagnostic for constraining the hydrogen density and column density of the BLR gas 
in LRDs (see Figure~\ref{fig:BB_BD}).

\item CLOUDY simulations further show that reproducing the large broad $\Ha$ equivalent widths seen in high-redshift AGN 
requires both a nearly unity covering fraction and high gas density (see Figure~\ref{fig:EW}).
A high covering fraction alone is insufficient; dense, optically thick BLR gas is needed to match the upper envelope of the observed EW distribution.

\end{itemize}

In summary, the unique Balmer-transition properties of LRDs point toward a BLR embedded in an extremely dense, possibly clumpy 
gaseous envelope or cocoon with an almost unity covering fraction.
Such conditions arise naturally in models of rapidly growing, newly born (seed) BHs \citep[e.g.,][]{Inayoshi&Ho2025}.
The combined constraints from Balmer-line ratios, Balmer-break strengths, and broad-line EWs therefore offer a direct glimpse 
into the earliest, gas-enshrouded AGN phase, possibly capturing the birth environments of the first generation of 
supermassive BHs.

\begin{acknowledgments}
We greatly thank Changhao Chen, Luis C. Ho, Roberto Maiolino, Jinyi Shangguan, and Benny Trakhtenbrot for constructive discussions.
K.I. acknowledges support from the National Natural Science Foundation of China (12573015, 1251101148, 12233001), 
the Beijing Natural Science Foundation (IS25003), and the China Manned Space Program (CMS-CSST-2025-A09).

This work is based on observations made with
the NASA/ESA/CSA James Webb Space Telescope.
The data were obtained from the Mikulski Archive for
Space Telescopes at the Space Telescope Science Institute, which is operated by the Association of Universities
for Research in Astronomy, Inc., under NASA contract
NAS 5-03127 for JWST. (Some of) The data products presented herein were retrieved from the Dawn JWST Archive (DJA). DJA is an initiative of the Cosmic Dawn Center (DAWN), which is funded by the Danish National Research Foundation under grant DNRF140.

\end{acknowledgments}

\facilities{JWST(NIRSpec; \citealt{Rigby23_JWST})}

\software{astropy \citep{2013A&A...558A..33A,2018AJ....156..123A,2022ApJ...935..167A},  
          Cloudy \citep{2023RNAAS...7..246G}, 
          dust\_extinction \citep{Gordon2024}
          }

\appendix

\begin{table*}[ht]
\caption{Source list of the LRDs used for the orange star symbols in Figure~\ref{fig:BB_BD}.
For each object, we list the survey field, object ID, sky coordinates, redshift, Balmer-break strength (BB), 
and broad-line Balmer decrement $\Ha/\Hb$ (BD).
For LRDs without published BB measurements, we computed the values directly from spectra reduced using 
the Dawn JWST Archive \citep{Heintz24_DJA}.}
\centering
\begin{tabular}{lllllllll}
\toprule
        Program-Field & ID & R.A. & Dec. & $z$ & BB & BD & Reference \\ \midrule
        JADES-GN & 68797 & 189.2291 & 62.1462 & 5.0405 & $1.726\pm0.135$ & $16.57^{+1.66}_{-1.62}$ & \cite{Nikopoulos2025} \\
        JADES-GN & 73488 & 189.1974 & 62.1772 & 4.1327 & $1.156\pm0.144$ & $19.64^{+3.16}_{-2.61}$ & \cite{Nikopoulos2025} \\
        RUBIES-EGS & 49140 & 214.8922 & 52.8774 & 6.6847 & $2.609\pm0.241$ & $8.72^{+0.37}_{-0.39}$ & \cite{Nikopoulos2025} \\
        CEERS-EGS & 1244 & 215.2406 & 53.0360 & 4.4771 & $1.307\pm0.154$ & $>6.83$ & \cite{Nikopoulos2025} \\
        JADES-GN & 53501 & 189.2951 & 62.1936 & 3.4294 & $1.337\pm0.269$ & $8.44^{+1.43}_{-1.21}$ & \cite{Nikopoulos2025} \\
        JADES-GN & 38147 & 189.2707 & 62.1484 & 5.8694 & $1.070\pm0.127$ & $>25.22$ & \cite{Nikopoulos2025} \\
        RUBIES-EGS & 50052 & 214.8234 & 52.8303 & 5.2393 & $1.099\pm0.126$ & $3.29^{+1.54}_{-0.91}$ & \cite{Nikopoulos2025} \\
        RUBIES-UDS & 154183 & 34.4107 & -5.1296 & 3.546 & $7.037\pm4.107$ & $16.21\pm8.54$ & \cite{deGraaff25} \\
        JADES-GS & 159717 & 53.0975 & -27.9012 & 5.0744 & $0.930\pm0.132$ & $18.5\pm2.7$ & \cite{DEugenio25} \\
        UNCOVER-A2744 & QSO1 & Multiple & Multiple & 7.0346 & $3.074\pm0.698$ & $7.472\pm0.378$ & \cite{Furtak25} \\
\bottomrule
\end{tabular}
\label{tab:BB_BD}
\end{table*}

\begin{table*}[ht]
\caption{
Balmer decrement $F(\mathrm{H}\alpha)/F(\mathrm{H}\beta)$ predicted by our CLOUDY simulations for a grid of hydrogen column densities $N_{\rm H}$ and hydrogen volume densities $n_{\rm H}$. 
}
\vspace{4mm}
\centering
\begin{tabular}{c*{11}c}
\toprule
& \multicolumn{11}{c}{$\mathrm{log}(n_{\rm H}/\mathrm{cm}^{-3})$} \\
$\mathrm{log}(N_{\rm H}/\mathrm{cm}^{-2})$ & 2.0 & 3.0 & 4.0 & 5.0 & 6.0 & 7.0 & 8.0 & 9.0 & 10.0 & 11.0 & 12.0 \\
\midrule
22.0 & 2.842 & 2.827 & 2.771 & 2.349 & 0.915 & 2.135 & 16.717 & 21.415 & 3.032 & 3.333 & 1.501 \\
22.5 & 2.854 & 2.781 & 2.380 & 1.175 & 2.868 & 19.743 & 49.261 & 16.107 & 6.520 & 3.728 & 2.552 \\
23.0 & 2.841 & 2.640 & 1.810 & 1.551 & 7.575 & 44.969 & 71.940 & 14.891 & 8.060 & 5.859 & 3.965 \\
23.5 & 2.826 & 2.464 & 1.442 & 2.836 & 13.564 & 73.734 & 73.348 & 13.250 & 7.234 & 6.846 & 4.579 \\
24.0 & 2.826 & 2.464 & 1.399 & 3.847 & 21.695 & 97.738 & 62.881 & 12.101 & 7.845 & 7.650 & 4.948 \\
\bottomrule
\end{tabular}
\label{tab:BD}
\end{table*}

\begin{table*}[ht]
\caption{
Balmer break strength predicted by our CLOUDY simulations for a grid of hydrogen column densities $N_{\rm H}$ and high hydrogen volume densities $n_{\rm H}$. 
For each pair of ($\nH,\NH$), three quantities are listed: (1) the Balmer break strength measured from the transmitted continuum 
($X=f_{\lambda_{\rm B,red}}^{\rm t}/f_{\lambda_{\rm B,blue}}^{\rm t}$), 
(2) the Balmer jump strength measured from the emitted continuum ($Y=f_{\lambda_{\rm B,red}}^{\rm e}/f_{\lambda_{\rm B,blue}}^{\rm e}$), and
(3) the flux density ratio at $\lambda_{\rm B,blue}=3600~{\rm \AA}$ between the transmitted and emitted continuum 
($Z=f_{\lambda_{\rm B,blue}}^{\rm e}/f_{\lambda_{\rm B,blue}}^{\rm t}$), respectively. 
Combining the covering factor $C$, the Balmer break strength of the observed spectrum is as ${\rm BD}=(X+YZC)/(1+ZC)$.
Values for $n_{\rm H} < 10^{7}\,\mathrm{cm^{-3}}$ are not shown because no significant Balmer break is produced at the lower densities.
}
\vspace{4mm}
\centering
\begin{tabular}{c*{6}c}
\toprule
& \multicolumn{6}{c}{$\mathrm{log}(n_{\rm H}/\mathrm{cm}^{-3})$} \\
$\mathrm{log}(N_{\rm H}/\mathrm{cm}^{-2})$ & 7.0 & 8.0 & 9.0 & 10.0 & 11.0 & 12.0 \\
\midrule
       
        ~ & 0.888 & 0.887 & 0.922 & 0.899 & 0.892 & 0.889 \\ 
        22.0 & 0.280 & 0.282 & 0.267 & 0.277 & 0.285 & 0.336 \\ 
        ~ & 0.104 & 0.101 & 0.116 & 0.152 & 0.147 & 0.123 \\ \hline
        ~ & 0.887 & 0.936 & 1.256 & 1.146 & 1.044 & 1.007 \\ 
        22.5 & 0.297 & 0.275 & 0.196 & 0.159 & 0.162 & 0.189 \\ 
        ~ & 0.116 & 0.133 & 0.28 & 0.494 & 0.797 & 0.712 \\ \hline
        ~ & 0.891 & 1.013 & 2.058 & 1.675 & 1.439 & 1.834 \\ 
        23.0 & 0.315 & 0.269 & 0.172 & 0.116 & 0.117 & 0.111 \\ 
        ~ & 0.130 & 0.182 & 0.707 & 1.259 & 1.785 & 3.791 \\ \hline
        ~ & 0.898 & 1.108 & 4.327 & 3.192 & 2.403 & 3.608 \\ 
        23.5 & 0.331 & 0.265 & 0.194 & 0.107 & 0.096 & 0.114 \\ 
        ~ & 0.145 & 0.244 & 1.778 & 3.263 & 4.284 & 8.967 \\ \hline
        ~ & 0.910 & 1.203 & 8.653 & 8.785 & 5.533 & 10.364 \\ 
        24.0 & 0.344 & 0.263 & 0.25 & 0.137 & 0.099 & 0.149 \\ 
        ~ & 0.163 & 0.313 & 3.412 & 8.611 & 11.351 & 23.873 \\ 
        \bottomrule
\end{tabular}
\label{tab:BB}
\end{table*}

\section{CLOUDY simulation data for Balmer breaks and decrements}\label{sec:appA}

Table~\ref{tab:BD} summarizes the Balmer decrements $F(\Ha)/F(\Hb)$ predicted by our CLOUDY simulations 
across a grid of hydrogen column densities $\NH$ and hydrogen volume densities $\nH$. 
These values serve as a direct diagnostic for interpreting observed Balmer-line ratios in dense gas 
and allow for the inference of the underlying physical conditions in high density environments.

Table~\ref{tab:BB} lists the Balmer break/jump properties computed for the same $(\NH,\nH)$ grid. 
For each model, we tabulate these quantities:\\
1. Balmer-break strength of the transmitted spectrum ($f_\lambda^{\rm t}$):
\begin{equation}
    X=f_{\lambda_{\rm B,red}}^{\rm t}/f_{\lambda_{\rm B,blue}}^{\rm t},
\end{equation}
2. Balmer jump strength of the emission spectrum ($f_\lambda^{\rm e}$):
\begin{equation}
    Y=f_{\lambda_{\rm B,red}}^{\rm e}/f_{\lambda_{\rm B,blue}}^{\rm e},
\end{equation}
3. Relative contribution of the emitted to transmitted continuum at $\lambda_{\rm B,blue}=3600~{\rm \AA}$:
\begin{equation}
    Z=f_{\lambda_{\rm B,blue}}^{\rm e}/f_{\lambda_{\rm B,blue}}^{\rm t}.
\end{equation}
These three quantities allow a direct computation of the Balmer break strength of the observed spectrum including the Balmer jump of 
the nebular continuum for any assumed geometric covering factor $C$. 
The combined observed Balmer break is
\begin{equation}
    {\rm BB}=\frac{X+YZC}{1+ZC}
\end{equation}
which accounts for the mixture of transmitted and emitted continua from a partially covered ionizing source.
Values for $\nH<10^7~\cc$ are omitted because the computed Balmer break is negligible in the low-density regime.

Taken together, Tables~\ref{tab:BD} and \ref{tab:BB} provide practical diagnostics for high-redshift AGNs including LRDs
(see Figure~\ref{fig:BB_BD}), linking gas density, column density, and the resulting optical/near-UV recombination signatures in high-density AGN environments.

\section{Balmer decrement in the LTE states}\label{sec:appB}

In the high density regime ($\nH\gtrsim10^{9}~\cc$), collisional excitation and de-excitation dominate radiative processes, driving the atomic level populations toward their LTE values. As a result, the Balmer decrement converges to an asymptotic value. 
This limiting value can be approximated as
\begin{equation}
    \frac{F(\Ha)}{F(\Hb)}\simeq\frac{\nu_{32}A_{32}}{\nu_{42}A_{42}}\left(\frac{n_3}{n_4}\right)_{\rm LTE}\frac{\beta_{32}}{\beta_{42}}
\end{equation}
where $A_{32}=4.410\times10^7~{\rm s^{-1}}$ and $A_{42}=8.419\times10^6~{\rm s^{-1}}$ are the Einstein coefficients averaged over 
angular momentum sub-levels, under the assumption of LTE within each principal quantum number for $n\geq3$ \citep{Janev1987,Omukai2001}. 
As shown in Figure~\ref{fig:tau_nH}, both $\Ha$ and $\Hb$ reach very large optical depth ($\tau_{\rm H\alpha}>\tau_{\rm H\beta}\simeq 10^4$) 
in this density regime and photon transport proceeds predominantly through resonance scattering within the Voigt profile. 
The corresponding escape probability scales as $\beta_{ul}\propto \tau_{ul}^{-1/2}$ \citep{Averett&Hummer1965,Avery&House1968,Kwan&Krolik81}, 
which yields
\begin{equation}
    \frac{\beta_{32}}{\beta_{42}}\simeq\sqrt{\frac{\tau_{42}}{\tau_{32}}}\simeq\sqrt{\frac{A_{42}\lambda_{42}^3g_4}{A_{32}\lambda_{32}^3g_3}}
\end{equation}
The LTE ratio of level populations is given by the Boltzmann distribution,
\begin{equation}
    \left(\frac{n_3}{n_4}\right)_{\rm LTE}=\frac{g_3}{g_4}\exp\left(\frac{h\nu_{34}}{k_{\rm B}T}\right)
\end{equation}
where $T$ is the gas temperature obtained from the CLOUDY simulations. 
Combining these pieces, the asymptotic Balmer decrement becomes
\begin{equation}
    \frac{F(\Ha)}{F(\Hb)}\simeq\sqrt{\frac{A_{32}g_3}{A_{42}g_4}}\left(\frac{\lambda_{42}}{\lambda_{32}}\right)^{5/2} 
    \exp\left(\frac{h\nu_{34}}{k_{\rm B}T}\right) \simeq 2.20 \cdot \exp \left(\frac{0.765}{T_4}-1\right)
    \label{eq:HaHb_LTE}
\end{equation}
where $T_4\equiv T/(10^4~{\rm K})$. 
The gas temperature in the neutral region of the gas slab depends weakly and negatively on the hydrogen column density
as $T\simeq 6000-8000~\K$ over $\NH=10^{22}-10^{24}~\Ncc$.
Therefore, the asymptotic Balmer decrement increases with $\NH$, consistent with the results shown in Figure~\ref{fig:SED_BD_nH}.

For the higher-order Balmer line ratio, Equation~(\ref{eq:HaHb_LTE}) can be generalized for $n\geq 4$ as 
\begin{equation}
    \frac{F(\Ha)}{F({\rm H}n)}\simeq\sqrt{\frac{A_{32}g_3}{A_{n2}g_n}}\left(\frac{\lambda_{n2}}{\lambda_{32}}\right)^{5/2} 
    \exp\left(\frac{h\nu_{3n}}{k_{\rm B}T}\right),
    \label{eq:HaHg_LTE}
\end{equation}
where the form is valid only when the optical depth of H$n$ is sufficiently high that $\beta_{n2}\propto \tau_{{\rm H}n}^{-1/2}$.
Because higher-order Balmer transitions generally have lower optical depths, this expression should be applied only 
within the appropriate physical regime.
For instance, one obtains $F(\Ha)/F(\Hg)\simeq 2.42 \cdot \exp \left(1.12/T_4-1\right)$.
The Balmer line ratios become $F(\Ha)/F(\Hb)=2.90$ and $F(\Ha)/F(\Hg)=5.75$ at $T=6000~\K$,
both of which are consistent with the Case~B values of $2.86$ and $6.10$ \citep{Osterbrock&Ferland2006}.

\bibliography{reference}{}
\bibliographystyle{aasjournalv7}

\end{document}